\documentclass[oldversion]{aa}

\usepackage{graphicx}
\usepackage{txfonts}
\usepackage{natbib}
\usepackage{supertabular}
\usepackage{longtable}
\usepackage{url}
\usepackage{dcolumn}

\begin{document}
\newcommand{\dla}{damped Lyman-$\alpha$}
\newcommand{\Dla}{Damped Lyman-$\alpha$}
\newcommand{\lya}{Ly-$\alpha$}
\newcommand{\lyb}{Ly-$\beta$}

%Title of paper
\title{HD molecules at high redshift \thanks{Based on observations carried out at the European Southern Observatory, under programme 278.A-5062 
with the Ultraviolet and Visual Echelle Spectrograph installed at the Very Large Telescope, unit Kueyen, 
on mount Paranal in Chile.}}%:
\subtitle{A low astration factor of deuterium in a solar-metallicity DLA system at $z$~=~2.418
}

\author{P. Noterdaeme\inst{1}, P. Petitjean\inst{1},  C. Ledoux\inst{2},  R. Srianand\inst{3} and A. Ivanchik\inst{4}}
\institute{Institut d'Astrophysique de Paris, CNRS - Universit\'e Pierre et Marie Curie, 98bis bd Arago, 75014 Paris, France\\
\email{noterdaeme@iap.fr, petitjean@iap.fr}
\and
European Southern Observatory, Alonso de C\'{o}rdova 3107, Casilla 19001, Vitacura, Santiago 19, Chile\\
\email{cledoux@eso.org}
\and
Inter University Centre for Astronomy and Astrophysics, Post Bag 4, Ganesh Khind, Pune 411\,007, India\\
\email{anand@iucaa.ernet.in}
\and
Ioffe Physical Technical Institute, St Petersburg 194021, Russia\\
\email{iav@astro.ioffe.ru}
}
\authorrunning{P. Noterdaeme et al.}
\titlerunning{HD at high-$z$: A low astration factor of deuterium in a solar metallicity DLA system at $z=2.418$}

\abstract{
We present the detection of deuterated molecular hydrogen (HD) in 
the remote Universe in a damped Lyman-$\alpha$ cloud at $z_{\rm abs}=2.418$ 
toward the quasar SDSS\,J143912.04$+$111740.5. This is a unique system in which 
H$_2$ and CO molecules are also detected. The chemical enrichment of this gas derived from Zn\,{\sc ii} and 
S\,{\sc ii} is as high as in the Sun. 
We measure $N({\rm HD})/2 N({\rm H_2})=1.5^{+0.6}_{-0.4}\times10^{-5}$, which is significantly higher than the same
ratio measured in the Galaxy and close to the primordial D/H ratio estimated from the WMAP constraint on the baryonic
 matter density ($\Omega_{\rm b}$). This indicates a low astration factor of deuterium that contrasts with the unusually
 high chemical enrichment of the gas. This can be interpreted as the consequence of an intense infall of primordial gas onto 
the associated galaxy.
Detection of HD molecules at high-$z$ also opens the possibility to obtain an independent 
constraint on the cosmological-time variability of $\mu$, the proton-to-electron mass ratio.
}

\keywords{Cosmology: observations -- 
          Galaxies: high-redshift -- 
          Galaxies: ISM --
          Quasars: absorption lines --
          Quasars: individual: SDSS J143912.04$+$111740.5}

\maketitle

\section{Introduction}

Deuterium is produced by primordial nucleosynthesis and is subsequently destroyed in stars. 
Therefore measurements of D/H from primordial gas provide important constraints on the
baryonic matter density $\Omega_{\rm b}$ in the framework of Big-Bang cosmology \citep{Wagoner73}.
Measurements at different redshifts in turn provide important clues on the star formation history 
\citep{Daigne04,Steigman07}. 
All the available D/H measurements at high-$z$ are based on the determination of the
$N($D$^0)/N($H$^0)$  ratio in low-metallicity QSO absorption line systems.
These measurements are difficult mainly because the velocity separation between
D\,{\sc i} and H\,{\sc i} absorption lines is small 
($\Delta v_{\ion{D}{i}/\ion{H}{i}}\sim80$~km\,s$^{-1}$) implying the lines are easily blended.
An additional difficulty is the presence of the Lyman-$\alpha$ forest, making 
hard to find the true continuum position and to discern between D\,{\sc i} absorption lines and 
intervening Ly-$\alpha$ forest lines. 
This explains why, despite more than a decade of efforts, only seven robust measurements of D/H at high-redshift have been performed \citep{OMeara06,Pettini08a}.  
It happens that these measurements are consistent with the
value [D/H]$_{\rm p}$ = 2.55$\pm 0.10\times10^{-5}$ derived from 
the Wilkinson Microwave Anisotropy Probe (WMAP) five year data 
\citep{Komatsu08} together with the $\eta\leftrightarrow$ [D/H]$_{\rm p}$ conversion from \citet{Burles01}, where 
$\eta$ is the baryon-to-photon ratio.
We consider this [D/H]$_{\rm p}$ value as the primordial D abundance in this work.
\par
Damped Lyman-$\alpha$ systems (DLAs) are absorbers with the highest
neutral hydrogen column densities among H\,{\sc i} absorption line systems 
($N($H$^0)\ga10^{20}$~cm$^{-2}$) and are thought 
to arise from the neutral interstellar medium (ISM) of distant galaxies \citep{Wolfe05}. 
Only a small fraction ($\sim$10-15\%) of them show detectable amounts of H$_2$ \citep{Ledoux03,Noterdaeme08}.
Among the 14 high-$z$ H$_2$-bearing DLAs known to date, only two show HD absorption 
\citep{Varshalovich01,Srianand08}. The $z_{\rm abs}=2.418$ system toward 
SDSS\,J143912$+$111740 we present here is the only one where, in
addition to H$_2$ and HD, CO is detected. 
From the excitation of CO it is possible to derive in a straightforward way 
an accurate estimate of the Cosmic Microwave Background Radiation temperature 
at the redshift of the absorber \citep{Srianand08}.
Here we focus on the HD/H$_2$ ratio and its implications for the star-formation history in the DLA galaxy.

\section{Observations  and Analysis\label{obs}}

Observations were performed with the Ultraviolet and Visual Echelle
Spectrograph of the Very Large Telescope at the European Southern
Observatory on March 21-25, 2007 with total exposure time on source exceeding 8\,h. Both blue and 
red spectroscopic arms were 
used simultaneously using dichroic settings with central wavelengths of resp. 390~nm and 580~nm (or 610~nm). 
The resulting wavelength coverage is 330$-$710~nm with a small gap between 452 and 478~nm.
The CCD pixels were binned 2$\times$2 and the slit width adjusted to $1^{\prime\prime}$ matching the seeing
conditions of $\sim 0.\!\!^{\prime\prime}9$. This yields a resolving power of $R=50\,000$, as measured on the 
thorium-argon lines from the calibration lamp. 
The data were reduced using the UVES pipeline v\,3.3.1 based on the ESO common pipeline library system. 
Accurate tracking of the object is achieved even in case of very low signal-to-noise ratio. 
Both the object and sky spectra are optimaly extracted and cosmic ray impacts and CCD defects are 
rejected iteratively.
Wavelengths were rebinned to the vacuum-heliocentric rest frame and individual scientific
exposures were co-added using a sliding window and weighting the signal by the total errors in each pixel. 
The dispersion around the wavelength calibration solution is 150~m\,s$^{-1}$.
Standard Voigt-profile fitting methods were used for the analysis to determine column densities, redshifts, and 
Doppler parameters~$b$. 

\begin{figure}%[!Ht]
\centering
\includegraphics[bb=85 90 570 730,angle=90,width=\hsize,clip=]{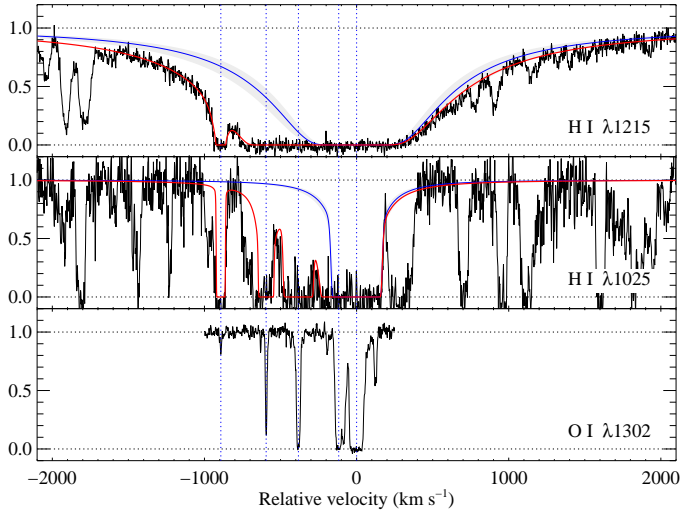}
\caption{Voigt profile fits to the Ly$\alpha$ (top panel) and Ly$\beta$ (middle panel) lines.
The bottom panel shows the absorption profile of O\,{\sc i} to visualize the positions
of individual H\,{\sc i} components (vertical dotted lines). The solid red
lines indicate the overall fit-model. Profiles are plotted on a velocity scale with 
origin set at $z_{\rm abs} = 2.41837$.
Blue profiles at $v=0$~km\,s$^{-1}$ represent the main H$^0$ component and the associated 
uncertainty as a shaded region. }
\label{HI} 
\end{figure}

H\,{\sc i}~absorption corresponding to the DLA is spread over $\sim$1000 km~s$^{-1}$. 
The velocity structure can be modeled using the asymmetry of the \lya\ line and 
the profile of the \lyb\ line (see Fig.~\ref{HI}). 
The derived structure is consistent with the velocity profile of the
O\,{\sc i} absorption. The latter species 
is believed to track H\,{\sc i} closely through charge exchange reactions 
for $\log N($H$^0)\ge19.0$ \citep{Viegas95}.
The continuum is fitted at the same time as the absorption lines with a large-scale low-order 
spline.

The column density in the main clump is strongly constrained by the red damping wing and the gaps
in the Ly-$\beta$ absorption.
The best fit gives $\log N($H$^0)=20.10^{+0.10}_{-0.08}$ for the main clump at $z_{\rm abs}$ = 2.41837.
Note that a column density of $\log N($H$^0)$~=~20.2 would fill in the whole red profile (see Fig.~1). 
For a column density smaller than $\log N($H$^0)$~=~20, the shape of the profile can not be reproduced 
by any absorption located at $-$117~km\,s$^{-1}$.

The overall fit is represented in red on Fig.~\ref{HI}, while the contribution from the main
clump only is shown by the blue profile and the associated uncertainties by the shaded region. 
Indicative column densities for other components are $\log N($H$^0)=$~17.9, 19.25, 19.20 and 19.20 at 
velocities $v$~=~$-$892, $-$594, $-$382, and $-$117~km\,s$^{-1}$, relative to the main clump
studied here. 
The complex velocity structure and/or insufficient signal-to-noise ratio in the blue prevent the determination of
$N($D$^0)$ in these individual components directly from D\,{\sc i} lines. 
From fitting together various optically thin transitions of S\,{\sc ii}, Zn\,{\sc ii}, Si\,{\sc ii} and
 Fe\,{\sc ii}, we derive metallicities in the main clump (at $v=0$~km\,s$^{-1}$; see Table~\ref{abundances}).
No ionisation correction is applied. The presence of strong neutral carbon and molecular lines in 
the main clump, including easily photo-dissociated CO \citep[][]{Srianand08} indicates that the effect 
of ionisation on abundances should be negligible. Indeed, in the abscence of neutral carbon, the ionisation 
correction for $\log N($H\,{\sc i}$)=20.1$ is smaller than 0.10~dex \citep[see Fig.~23 of][]{Peroux07}.
The S and Zn metallicities are solar and the relative depletion 
pattern (from Si and Fe) is typical of what is seen in cold neutral ISM 
clouds in the Galaxy (see Table~\ref{abundances}). 

%ABUNDANCES
\begin{table}
\centering
\caption{\label{abundances} Observed metal abundances}
\begin{tabular}{c c c}
\hline
\hline
Ion (X)   & $\log N$(X)    & [X/H]$^1$ \\
\hline                                                             %with M03           
S$^+$      & $15.27\pm0.06$                                       & $-0.03\pm0.12$   \\
Zn$^+$     & $12.93\pm0.04$                                       & $+0.16\pm0.11$   \\
Si$^+$     & $14.80\pm0.04$                                       & $-0.86\pm0.11$   \\
Fe$^+$     & $14.28\pm0.05$                                       & $-1.32\pm0.11$   \\
N$^0$      & $\ge15.71$                                           & $\ge-0.34$       \\
\hline
\end{tabular}\\
\footnotesize
$^1$ with respect to solar abundances from \citet{Morton03}.
\normalsize
\end{table}
 
\section{Column densities of molecules and D/H ratio \label{dh}}

\begin{figure}
\centering
\includegraphics[bb=85 90 570 730,angle=90,width=\hsize,clip=]{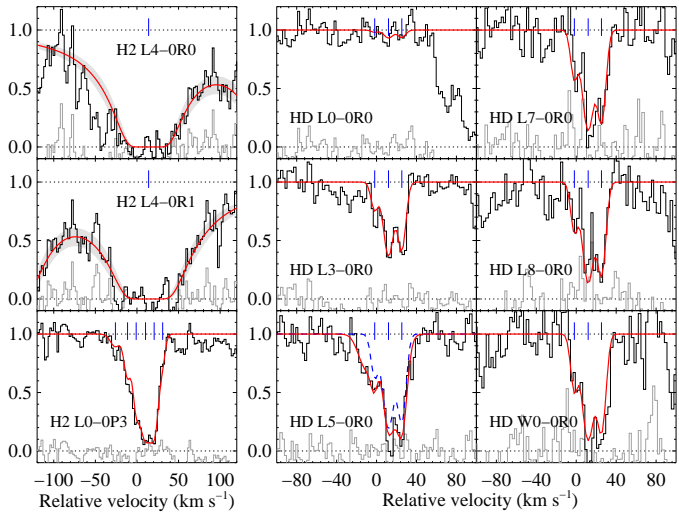}
\caption{\label{fits} Molecular hydrogen absorption lines. 
The normalized flux is given on a velocity scale with the origin set at
 $z_{\rm abs}=2.41837$. The Voigt profile fits as well as the residuals are also plotted. 
The blue dashed profile overplotted on HD\,L5-0\,R0 corresponds to the contribution from HD only. 
The grey regions for H$_2$\,L4-0\,R0 and H$_2$\,L4-0\,R1 represent the 1$\sigma$ uncertainty around 
the best fit. Short vertical lines mark the position of individual components.}
\end{figure}

%H2
Absorption lines from more than one hundred H$_2$ transitions
from rotational levels J~=~0 up to J~=~5 are detected in the main H\,{\sc i} component
in six components spread over 50 km~s$^{-1}$. Column densities in different J levels 
are obtained by simultaneous fits and are especially well constrained by the presence 
of damping wings seen on the corresponding lines from the Lyman 2-0, 4-0, 5-0, 7-0, 
8-0, 9-0, and 10-0 bands and from the Werner 0-0 and 1-0 bands. Examples are shown on Fig.~\ref{fits}, 
with $\chi^2$~=~1.04, 0.99 and 0.98 for J~=~0, 1 and 3, respectively.
We measure a total column density of $\log N({\rm H_2})=19.38\pm0.10$ and
a molecular fraction, $f=2N({\rm H_2})/[2N({\rm H_2})+N($H$^0)]=0.27^{+0.10}_{-0.08}$.
The uncertainty on $N($H$_2$) is dominated by the error from fitting the 
damping wings of low rotational level lines (J~=~0 and 1).

%HD
Deuterated molecular hydrogen is detected in the first rotational level in three components 
associated to the strongest H$_2$ components.  Five HD absorption lines 
(L3-0\,R0, L5-0\,R0, L7-0\,R0, L8-0\,R0 and W0-0\,R0) are clearly detected 
and were fitted simultaneously. The fits are shown on Fig.~\ref{fits}. Signal-to-noise 
ratios are $\sim$12 for L0-0\,R0, L3-0\,R0 and L5-0\,R0, and $\sim$6 for the remaining lines.
$\chi^2$ values for the fit to these lines are 0.43 (L0-0\,R0), 1.04 (L3-0\,R0), 
 1.36 (L7-0\,R0), 1.07 (L8-0\,R0) and 1.49 (W0-0\,R0). 
Note that fittings were done independently by two of us (PN and RS) with two different 
tools (FITLYMAN and VPFIT), yielding same results.
The strongest constraint comes from L3-0\,R0 transition which has good SNR and is optically thin (see Fig.~2). 
Note that L5-0\,R0 is blended with another absorption and is not included in the fit.
Results of the fits are summarized in Table~\ref{HDtab}. Errors correspond the fit 
of both the lines and the continuum. 
The total HD column density is $\log N({\rm HD})=14.87\pm0.025$. 

\begin{table}%[!Ht]
\centering
\caption{\label{HDtab} Voigt profile fitting results for HD} 
\begin{tabular}{c c c c}
\hline
\hline
$z_{\rm abs}$ & $v^1$ & $b$ & $\log N$(HD) \\
             & (km\,s$^{-1}$) &  (km\,s$^{-1}$) &   \\
\hline
2.41835 &    $-$1.9       & $2.9\pm0.8$       & $13.89\pm0.08$\\
2.41851 &    12.0         & $5.0\pm1.0$       & $14.57\pm0.04$\\
2.41866 &    25.3         & $3.5\pm0.3$       & $14.46\pm0.03$\\
\hline
\end{tabular}\\
\footnotesize
$^1$ Velocity relative to $z_{\rm abs}=2.41837$.
\normalsize
\end{table}

\begin{figure}[!t]
\includegraphics[width=\hsize,bb=23 410 600 750]{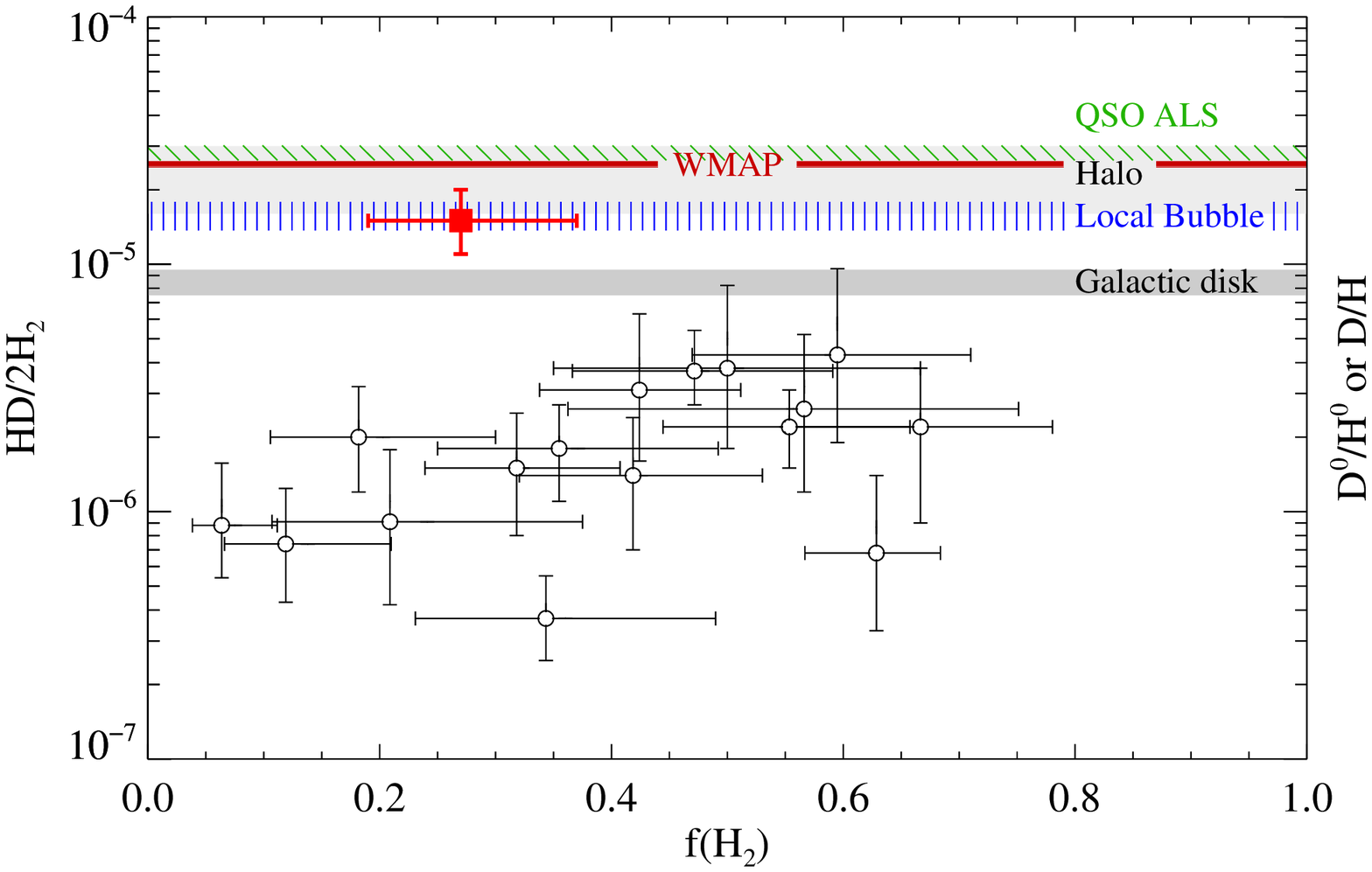}
\caption{\label{HDH2} The {\bf HD/2H$_2$} ratio vs the molecular fraction. 
The filled square is the new measurement at
 $z_{\rm abs}=2.418$ toward SDSS\,J143912$+$111740. 
The empty circles are {\sc fuse} and Copernicus measurements in
 the Galactic ISM \citep{Lacour05b}. The grey region
 marks the D$^0$/H$^0$ ratio in the Galactic disk \citep{Linsky06}, 
the light grey region marks this ratio in the Galactic halo 
\citep{Savage07} and the vertically-dashed one that in the Local Bubble \citep{Moos02,Linsky06}. 
The oblique dashed lines stand for D$^0$/H$^0$ measurements in high-redshift quasar absorption-line systems 
\citep[QSO ALS;][]{Pettini08a}. 
Finally, the solid line corresponds to the D/H ratio estimated from the baryon/photon ratio \citep[WMAP;][]{Komatsu08}.
}
\end{figure}

Although each HD component is associated to one of the H$_2$ components, the strong blending of the latter, 
especially in low rotational levels, does not allow for the determination of $N($HD$)/2N($H$_2)$ 
in individual components. 
We therefore use the column densities integrated over the whole profile for both HD and H$_2$ 
and obtain $N({\rm HD})/2N({\rm H_2})=1.5^{+0.6}_{-0.4}\times10^{-5}$.
In Fig.~\ref{HDH2}, we compare the $N($HD$)/2N($H$_2)$ ratio with similar measurements 
in the Galactic ISM \citep{Lacour05b}. 
It is apparent that the $N({\rm HD})/2N({\rm H_2})$ ratio in
the present system is an order of magnitude higher than the values for similar 
molecular fraction (and $N({\rm H_2})$) measured in the Galaxy.
We also compare these results to D$^0$/H$^0$ measurements in low metallicity 
clouds toward high-redshift quasars \citep{Pettini08a},  
the Galactic disk, the Galactic halo and the primordial D/H ratio estimated from the
five-year WMAP results \citep{Komatsu08}. 

From the molecular ratio, HD/2H$_2$, we derive 
(D/H)$_{DLA}>0.7\times10^{-5}$ at the 95\% confidence level. This 
corresponds to an astration factor $f_{\rm D}=({\rm D/H})_{\rm p}/{(\rm D/H)_{\rm DLA}}<3.6$
where (D/H)$_{\rm p}$ refers to the primordial abundance derived from WMAP. 
However, the true (D/H) ratio (resp. astration factor) is probably well above the
derived lower limit (resp. well below the upper limit) for various reasons:

First, the quoted HD/2H$_2$ ratio could represent a lower limit on the 
actual ratio in individual components as H$_2$ and HD are possibly 
not co-spatial. Indeed, the maximum HD column density does 
not arise in the strongest CO component \citep[see][]{Srianand08}, suggesting a 
fraction of the molecular hydrogen is not associated with HD.

Second, deriving the deuterium abundance from the HD column density is difficult 
because of the complex chemistry \citep[e.g.][]{Cazaux08} and the sensitivity of the HD 
abundance on the particle density, cosmic ray density and UV field \citep{LePetit02}.
However FUSE and Copernicus observations have shown that in the ISM of the Galaxy, the HD/H$_2$ ratio 
increases with the molecular fraction and that HD/2H$_2$ could trace D/H well only when 
$f\sim1$ \citep{Lacour05b}, when both HD and H$_2$ are self-shielded from photo-dissociation. 
In diffuse gas, the HD optical depth is expected to be smaller than that of H$_2$ --as
the deuterium abundance is low-- and HD/2H$_2$ will provide a lower limit on D/H 
\citep{LePetit02,Liszt06}. This is also supported by recent FUSE observations by \citet{Snow08}.

Finally, the D/H ratio in the gas phase, as measured by HD/2H$_2$, could itself be a lower limit on the 
true abundance ratio if depletion on dust is significant \citep{Prochaska05hd,Draine06}.

\section{Discussion}
\subsection*{Evidence for infall?}
The high D/H ratio inferred above indicates that astration of deuterium
is low eventhough metallicity is solar. This situation is well explained
in our Galaxy by models including infall of primordial material \citep{Steigman07,Romano06,Prodanovic08}.

If we use the sulfur abundance as a proxy for that of oxygen,
we note that the [N/O] ratio in the main H\,{\sc i} component
is consistent with the ratio expected for secondary nitrogen production at 
solar metallicity \citep{Centurion03,Petitjean08,Pettini08b}. In addition, the [S/Zn]
ratio is consistent with the solar value and does not indicate any $\alpha$-enhancement.

The inferred lower limit on D/H, the near solar value of [$\alpha$/Fe] and
[N/$\alpha$] rule out rapid star formation that is generally invoked to explain
high chemical enrichment in elliptical galaxies.

Speculating slightly further, we note that the metal profile is spread over at least $800$~km\,s$^{-1}$ (see Fig.~\ref{HI}).
This, together with the solar metallicity, is consistent with the gas being associated with 
a deep potential well \citep{Ledoux06a}.

The properties of the gas in the present system are similar to those of
the ISM in our Galaxy.
But these properties are reached on a time-scale five times smaller for the DLA galaxy than for the
Milky Way, as the age of the Universe at $z = 2.42$ is 20\% of its present age adopting the most recent 
cosmological parameters \citep{Komatsu08}. This must imply that the system has 
undergone continuous star-formation and infall. 

If one assumes negligible production of deuterium by cosmic rays and that this element  
is completely destroyed 
in the material that goes through stars then the gas infall rate ${\rm \dot M_{in}}$ should on 
average be of the order of the star formation rate ${\rm \dot M_{SFR}}$
in order to replenish deuterium. 
Recently several observational evidences have been published 
for cold gas accretion onto massive galaxies
at high-$z$ \citep{Weidinger05,Nilsson06,Dijkstra06} 
and low-$z$ \citep{Fraternali08}. 
In addition, numerical simulations suggest that at $z= 2-3$ the accretion of cold material from
the IGM dominates for halos with masses $<\sim3\times10^{11}$ M$_\odot$ \citep{Keres05}. Interestingly, 
the inference that ${\rm \dot M_{in}}\sim {\rm \dot M_{SFR}}$ is also required to understand 
the properties of $z = 2-3$ Lyman-break galaxies \citep{Erb08}.  
Our observations reinforce this important finding.

\subsection*{Constraint on the variability of $\mu$}

The detection of several HD transitions should make it possible to test the 
time variation of $\mu=m_p/m_e$, the proton-to-electron mass ratio. 
H$_2$ transitions at high redshift have been used \citep[e.g.,][]{Ivanchik05,Reinhold06,King08} 
to probe the variability of~$\mu$. 
This is done by measuring the relative position of the lines around the overall 
redshift of the absorbing cloud,
\[
\zeta_i={(z_i-\overline{z}_{\rm abs})/(1+\overline{z}_{\rm abs})}={{\Delta \mu} \over \mu} K_i~ ,
\]
where $z_i=\lambda_i/\lambda_i^0-1$ is the observed redshift of line $i$ and 
$K_i=d \ln \lambda_{i}^0/ d \ln \mu$ is the sensitivity on $\mu$ calculated for each transition. 
$\overline{z}_{\rm abs}$ is taken for each component as the weighted mean redshift from the 
different transitions. 
As these measurements may involve various
unknown systematics, it is important to use different sets of lines
and different techniques. Sensitivity coefficients and accurate wavelengths for HD transitions 
have been published very recently \citep{Ivanov08}. 
However, the SNR of our data at the position of the HD absorption lines 
prevents us to perform a measurement at the level of current studies. 
Additional data on this quasar are needed to be able to derive an independent
constraint on the variability of $\mu$.

\section{Conclusion}
We reported the detection of HD molecules at $z_{\rm abs}=2.418$ toward 
SDSS\,J143912$+$111740, following a careful selection of quasars in the 
SDSS database and intensive observations with UVES at the Very Large Telescope.
The system presents characteristics very similar to what is observed in the solar 
neighbourhood. We find HD/2H$_2=1.5\times10^{-5}$, consistent with an astration factor 
of deuterium less than 1.7 which is contrasting with the high chemical enrichment.
This is best explained by a scenario in which the gas that goes through star formation 
is replenished by continuous infall of ambient primordial gas. 
Similar results arise from recent numerical simulations 
and semi-analytical models \citep[e.g.][]{Keres05,Erb08}. 
Interestingly, dynamical studies of nearby galaxies \citep{Fraternali08} as 
well as interpretation of Ly-$\alpha$ blobs at high redshift \citep{Weidinger05,Nilsson06,Dijkstra06} 
provide independent observational evidences for accretion of gas onto massive galaxies.

We finally stress the importance of detecting similar systems 
to probe the time-variation of the proton-to-electron mass ratio from HD lines. 
 Although such a independent test would be welcome to characterize possible unknown systematics, 
significantly higher signal-to-noise ratio is required to obtain limits comparable to those obtained 
with other techniques.

\acknowledgement{We thank an anonymous referee for comments and suggestions that 
improved the quality of the paper.
We warmly thank the ESO Director Discretionary Time allocation committee 
and the ESO Director General, Catherine Cesarsky, for allowing us to carry out these observations.
PPJ and RS gratefully acknowledge support from the Indo-French Centre for the Promotion of
Advanced Research (Centre Franco-Indien pour la Promotion de la Recherche Avanc\'ee).}

\bibliographystyle{/scisoft/share/texmf/aa/aa-package/bibtex/aa}
\bibliography{hdbib}
\end{document}